\documentclass[sigconf]{acmart}
\acmConference{}{}{}

\AtBeginDocument{%
  }

\usepackage{pifont}

\begin{document}

\title{Privacy in Responsible AI: Approaches to Facial Recognition from Cloud Providers
}

\author{Anna Elivanova}
\affiliation{
\institution{University of Hull}
\city{Kingston-upon-Hull}
\country{United Kingdom}
}

\renewcommand{\shortauthors}{Elivanova}

\begin{abstract}
As the use of facial recognition technology is expanding in different domains, ensuring its responsible use is gaining more importance. This paper conducts a comprehensive literature review of existing studies on facial recognition technology from the perspective of privacy, which is one of the key Responsible AI principles. 

Cloud providers, such as Microsoft, AWS, and Google, are at the forefront of delivering facial-related technology services, but their approaches to responsible use of these technologies vary significantly. This paper compares how these cloud giants implement the privacy principle into their facial recognition and detection services. By analysing their approaches, it identifies both common practices and notable differences. The results of this research will be valuable for developers and businesses by providing them insights into best practices of three major companies for integration responsible AI, particularly privacy, into their cloud-based facial recognition technologies. 

\end{abstract}



\maketitle

\section{Introduction}

Recent developments in AI have brought significant changes to various domains of our lives. AI-based facial recognition technologies (FRT) have emerged as one of the most impactful but at the same time controversial applications. While these systems can be applied to a variety of scenarios in everyday life, like event security, travel security, and bank authentication, they also present significant challenges, particularly in privacy perspective. Privacy concerns become even more pronounced when the FRT system is deployed in the cloud, where sensitive data is processed and stored by third-party providers. 

The \textbf{goal} of this paper is to analyse how privacy is addressed in facial recognition services by major cloud providers - Microsoft Azure, Amazon Web Services (AWS), and Google Cloud - within the framework of Responsible AI. 

This paper is structured around the following \textbf{research questions}: 
\begin{enumerate}
    \item What are general definitions of the terms "Responsible AI" and "privacy" within its framework?
    \item What are the current privacy concerns and solutions in the context of FRT?
    \item Who is responsible for ensuring privacy in cloud-based facial recognition systems?
    \item   What measures do major cloud providers (Microsoft, AWS, Google) implement to support privacy in their facial recognition applications?
\end{enumerate}

This paper focuses on exploring privacy considerations in cloud-based facial recognition systems within the broader framework of Responsible AI. Specifically, it has the following \textbf{scope}:

\begin{itemize}
    \item   It investigates the current state of privacy measures and challenges in cloud-based facial recognition systems.
    \item The analysis is scoped to major cloud providers (Microsoft Azure, AWS, and Google Cloud) and their privacy measures related to facial recognition.
\end{itemize}

The study is subject to the following \textbf{limitations}:

\begin{itemize}
\item While the study may reference regulatory frameworks, it does not provide an in-depth jurisdiction-specific legal analysis due to the vast diversity of global privacy laws, which fall outside the scope of this research. 
\item The study does not address how customers of cloud providers address privacy in their implementation of facial recognition systems, as the focus is on the privacy approaches adopted by the cloud providers themselves. 
\end{itemize}

The paper begins with an introduction to Responsible AI, its principles, and privacy concerns in FRT. It then examines cloud-based facial recognition systems and outlines the methodology used. The discussion focuses on the shared responsibility model for data protection, followed by an analysis of privacy approaches by Microsoft, AWS, and Google. Each provider's privacy measures and guidelines are compared, and the paper concludes with a comparison of approaches and final conclusions.

\section{Background}

The systematic literature review for this study is based on the sources outlined in Table \ref{tab:sources_count}. The review is predominantly limited to peer-reviewed journal papers. However, certain exclusions were made, such as industry guidelines for AI principles, due to their importance and the fact that many other systematic academic reviews in the relevant fields already refer to these. Additionally, white papers from reputable research institutions and a widely cited book were also considered.

All sources included in the review were published within the past five years to ensure the relevance and timeliness of the analysis.

\begin{table}
    \centering
    \includegraphics[width=0.5\textwidth]{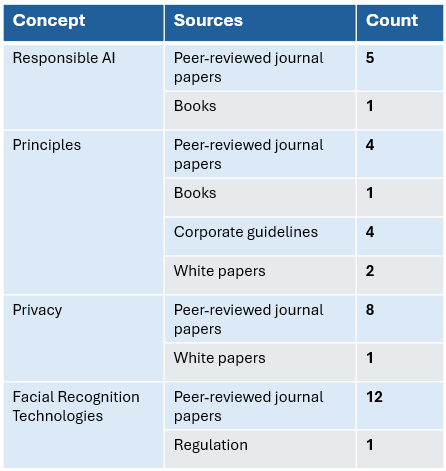}
    \caption{Distribution of sources by concept}
    \label{tab:sources_count}
\end{table}

\subsection{Responsible AI: Definitions and Core Principles} 
\subsubsection{Responsible AI}

The field of responsible AI has emerged and developed in response to the increasing prominence of AI systems in society. Despite this rapid development, there is no common approach to the definition of Responsible AI. Diverse approaches have been offered:

\begin{itemize}
    \item \textbf{Dignum (2019)} takes a philosophical approach, emphasizing the responsibility that comes with the power of AI \cite{dignum2019}.
\end{itemize}
\begin{itemize}
    \item \textbf{Goellner (2024)} defines Responsible AI as a human-centered approach that fosters user trust by ensuring ethical decision-making. According to the author, it also ensures that decisions are explainable and user privacy is protected through secure implementations \cite{goellner2024responsible}.
\end{itemize}
\begin{itemize}
    \item \textbf{Lu et al. (2024)} refer to Responsible AI as ethical development for societal benefit, emphasizing that AI should be designed not only for the benefit of individuals, but also society and the environment \cite{Lu}. This definition is less about the decision-making processes and more about the intent behind AI systems.
\end{itemize}
\begin{itemize}
    \item \textbf{Diaz-Rodriguez (2024)} offers a broader, systemic view, defining Responsible AI as the integration of key aspects of trustworthy AI, including ethical principles, philosophical approaches, regulatory frameworks, and technical requirements \cite{DIAZRODRIGUEZ}.
\end{itemize}.

Multiple sources highlight that Responsible AI is human-centred \cite{goellner2024responsible, dignum2019}

\subsubsection{Principles of Responsible AI}

The principles of Responsible AI are not legally binding obligations, but rather serve as guidelines that reflect moral obligations. However, there is no governing authority with the power to enforce these principles \cite{CLARKE2019}.

The most common principles in the reviewed frameworks are highlighted in Table \ref{tab:principles}.

\begin{table*}
  \caption{Key Principles Identified in Responsible AI Frameworks}
  \label{tab:principles}
  \begin{tabular}{lcccccccccc}
    \toprule
    Principles&  Clarke\cite{CLARKE2019}  &Dignum\cite{dignum2019}&Mikalef\cite{mikalef} &Liu\cite{Liu2023}  &Cheng \cite{Cheng}&Fjeld\cite{fjeld2020principled}&Microsoft\cite{microsoft_ai_principles} &AWS\cite{aws_responsible_ai}&IBM\cite{ibm_responsible_ai} &Google\cite{google_ai_principles} \\
    \midrule
    Fairness&    &\ding{51} &\ding{51} &\ding{51}   &\ding{51}   &\ding{51}  &\ding{51} &\ding{51} & \ding{51} &\ding{51}  \\
 Transparency& \ding{51}  &\ding{51} & \ding{51} & \ding{51}   &\ding{51}   & \ding{51} & \ding{51} & & \ding{51} & \\
    Privacy&     &\ding{51} & &\ding{51}   &&\ding{51} &\ding{51} &\ding{51} & \ding{51} &\ding{51}  \\
 Accountability& \ding{51}  &\ding{51} & \ding{51} &  && \ding{51}  & \ding{51} & & & \ding{51}  \\
 Safety&  &\ding{51} & \ding{51} &  &\ding{51}   & \ding{51} & \ding{51} & \ding{51} & & \ding{51}  \\
 Robustness& \ding{51}  && \ding{51} &  && & & \ding{51} & \ding{51} & \\
 Human Control& \ding{51}  && \ding{51} & \ding{51}   && \ding{51} & & \ding{51} & & \\
 Security&  &\ding{51} & &  \ding{51}   &&\ding{51} & \ding{51} & & & \\
 Explainability&  &\ding{51} & &  && \ding{51} & & \ding{51} & \ding{51} & \\
    Inclusiveness&     && &\ding{51}   &&&\ding{51} &&  & \\
 Societal Wellbeing&  && \ding{51} &  && & & & & \ding{51} \\
 Human Values& \ding{51}  && & \ding{51}   && \ding{51} & & & & \\
 Human Wellbeing& \ding{51}  && &   &&& & & & \\
   \bottomrule
\end{tabular}

\end{table*}

According to the results of the review, Fairness, Privacy and Transparency are the most emphasized principles, highlighted by nearly all sources, followed by Accountability, Safety, and Robustness. 

Notably, human-centric principles such as Human Wellbeing, Societal Wellbeing, and Human Values are among the least emphasized and not mentioned by commercial entities like Microsoft, AWS, and Google. This shows a divergence between theoretical frameworks and industry-driven ones. 

It was also observed that Microsoft and AWS treat privacy and security as a unified principle.  In the present research, security will be considered an integral part of privacy, as the two concepts are closely interconnected, and safeguarding data is inseparable from ensuring privacy.

\subsubsection{Privacy}

One of the mostly widely accepted general definitions of \textit{privacy} originates from Westin's "Privacy and Freedom" (1967) and refers to an individual's or group's ability to "determine for themselves when, how, and to what extent information about themselves is communicated to others" \cite{DOMINGOFERRER201938}. 

In the context of Responsible AI framework, privacy principle ensures that AI systems protect individuals' autonomy by giving them control over their data and decisions influenced by AI \cite{fjeld2020principled}.

Privacy considerations must span across the entire lifecycle of an AI system: 
\begin{itemize}
    \item Design and implementation. Diaz-Rodriguez argues that privacy should be embedded during the system design through robust data governance mechanisms, ensuring data quality and legitimate access \cite{DIAZRODRIGUEZ}. 
    \item Training and testing. AI models rely heavily on data during training, requiring methods to protect information during this phase. Technologies like federated learning are crucial here \cite{DIAZRODRIGUEZ, Yang2021}.
    \item Deployment and Operation. As Cheng (2020) notes, operational AI systems need mechanisms like encryption and regular audits to safeguard ongoing data usage \cite{Cheng}. 
\end{itemize}

A range of technologies has been developed to protect privacy:
\begin{itemize}
    \item Differential Privacy: This technique introduces noise into datasets to obscure individual contributions \cite{goellner2024responsible, FERDOWSI2024, Yang2021, Cheng, Okon2024}
    \item Federated Learning: Yang (2021) and Diaz-Rodriguez et al. (2020) emphasize the decentralized nature of federated learning, which enables collaborative AI model training without sharing raw data \cite{Yang2021, DIAZRODRIGUEZ, Okon2024}.
    \item Encryption techniques, including Homomorphic Encryption, Somewhat and Fully Homomorphic Encryption, Secure Multiparty Computation: These cryptographic methods allow computations on encrypted data, ensuring that data remains private during processing \cite{goellner2024responsible, Yang2021, DIAZRODRIGUEZ, Cheng, Okon2024, DOMINGOFERRER201938, Sivan2021}.
    \item Trusted Execution Environment: This technique provides a secure environment for data processing, encrypting data during transmission and minimizing risks \cite{Yang2021}. 
    \item     Data splitting mechanisms, i.e. fragmenting sensitive data and storing them in separate locations \cite{DOMINGOFERRER201938}. 
    \item Data anonymation methods, which consist in masking irreversibly data to preserve privacy \cite{DOMINGOFERRER201938}. 
\end{itemize}

Okon et al. (2024) emphasize the importance of privacy impact assessments to evaluate the risks associated with data processing. These assessments are especially important for public cloud environments, which present higher risks compared to private or on-premises environments \cite{Okon2024}. 

The following challenges and trades-off present privacy preservation:
\begin{itemize}
    \item Accuracy vs Privacy: Ferdowsi et al. (2024) note that while stronger privacy protections, like differential privacy, enhance security, they can reduce model accuracy \cite{FERDOWSI2024}.
    \item Efficiency vs Security: Advanced privacy enhancing technologies, such as secure multiparty computation, are computationally intensive, constituting challenges for practical implementation \cite{Yang2021}. 
\end{itemize}

\subsection{Facial Recognition Technologies}
\subsubsection{Responsible AI in Facial Recognition}

FRT is a powerful tool for identifying or verifying individuals by analysing facial features in digital images or videos. The technology operates through a sequential process comprising face detection, alignment, feature extraction, and matching \cite{YangCloud2021}. 

FRT can be categorized into distinct types based on its functionality and use cases. Table \ref{tab:frt_types}  summarises the primary types, their descriptions, alternative terminologies, and relevant references. 

\begin{table*}[h!] 
  \caption{Facial Recognition Technology types}
  \label{tab:frt_types}
  \centering
  \renewcommand{\arraystretch}{1.2} 
  \begin{tabular}{p{3cm} p{4cm} p{3cm} p{3cm} p{2cm}}
    \toprule
    \bfseries Type of FRT&  \bfseries Description& \bfseries Alternative Terminology& \bfseries Key Applications& \bfseries References\\
    \midrule
    Verification FRT&    Determines whether two images belong to the same person.&Authentication, One-to-One&Access control, device unlocking&\cite{YangCloud2021}, \cite{Matulionyte2024}, \cite{Buyukcavus2023}\\
 Identification FRT& Compares an image to a database to find a match&One-to-Many& Surveillance, criminal investigation& \cite{YangCloud2021}, \cite{Matulionyte2024}, \cite{Buyukcavus2023}, \cite{Nwafor2023}\\
    Categorization FRT&     Extracts characteristics, such as ethnicity, sex, or age from facial data&-& Demographic analysis, marketing&\cite{Matulionyte2024}\\
 Emotional FRT& Detects and interprets human emotions based on face expressions&-& Customer service, mental health&  \cite{Matulionyte2024}\\
\bottomrule
\end{tabular}

\end{table*}

In its turn, identification FRT is further classified into two categories, as outlined in Table \ref{tab:identification_frt}. 

\begin{table*}[h!] 
  \caption{Identification FRT Categories}
  \label{tab:identification_frt}
  \centering
  \renewcommand{\arraystretch}{1.2} 
  \begin{tabular}{p{3cm} p{4cm} p{3cm} p{3cm} p{2cm}}
    \toprule
    \bfseries Type of FRT&  \bfseries Description& \bfseries Alternative Terminology& \bfseries Key Applications& \bfseries References\\
    \midrule
    Ex-Post Identification&    Retrospective analysis of previously collected data&-&Post-event investigation&\cite{Matulionyte2024}\\
 Live Identification& Real-time identification from live camera feeds&-& Border control, even security& \cite{Matulionyte2024}\\
    \bottomrule 
\end{tabular}

\end{table*}
Although the applications of FRT are increasingly widespread, challenges associated with its responsible use have triggered significant debates. 

\subsubsection{FRT and Privacy}

Privacy is a critical concern in the development of FRT. The technology involves the collection and processing of sensitive biometric data, raising concern about unauthorized access, data misuse, and potential surveillance \cite{Nwafor2023}. Xu et al. demonstrated that perceived privacy concerns significantly influence public trust in FRT systems \cite{Xu2021}. When privacy is prioritized over security benefits, trust in these systems tend to increase. Conversely, neglecting privacy can erode public confidence and lead to resistance against FRT adoption \cite{Hyesun2024}. 

Several approaches have been proposed to address privacy concerns in FRT, combining regulatory frameworks with technical innovations:
\begin{itemize}
    \item Hyesun emphasized the responsibility of governments and institutions to implement policies that ensure privacy protection \cite{Hyesun2024}.
    \item The European Union's AI Act addresses privacy concerns by prohibiting practices such as unauthorized scraping of facial images, aiming to reduce mass surveillance and safeguard privacy rights \cite{eu_ai_act}. 
    \item Butt et al. proposed using task-specific, low resolution sensors to avoid capturing surrounding objects \cite{Butt2023} .
\end{itemize}

\subsubsection{Cloud-based FRT systems}
The integration of FRT with cloud computing introduces unique opportunities. Cloud-based FRT systems facilitate remote biometric identification by comparing input images with databases hosted in the cloud. These systems also enable real-time analysis, making them suitable for large-scale applications \cite{Buyukcavus2023}. Buyukcavus et al. evaluated the performance of  some cloud-based FRT APIs (namely, Microsoft Azure, AWS, and Face++), demonstrating their high accuracy even in challenging scenarios, such as identifying individuals who had undergone facial alterations \cite{Buyukcavus2023}. Furthermore, Yang et al. highlight the scalability of Cloud-based FRT, which allows its adaptability to diverse spatial architectures and its ability to scale according the varying demand \cite{YangCloudFRT}. 

Despite its advantages, cloud-based FRT introduce significant challenges:
\begin{itemize}
    \item Privacy risks: Sensitive biometric data stored in the cloud is vulnerable to breaches and unauthorized access \cite{HossainAnaMuhammad}. 
    \item Computational overhead: Advanced encryption techniques, such as homomorphic encryption and secure multiparty computation, enhance data security but remain computationally expensive \cite{YangCloud2021}. 
\end{itemize}

The general data protection methods discussed discussed in the \textit{Privacy} subsection are also applicable to cloud systems, specifically FRT. Additionally, specific measures are suggested to cloud environments, including: 
\begin{itemize}
    \item Access control \cite{DOMINGOFERRER201938, Sivan2021}
    \item Storage Control \cite{Sivan2021}
\end{itemize}

Cloud-based FRT providers have been criticised for ethical shortcomings in their implementation methods and data managements practices. Wen and Holweg documented the responses of companies like IBM, Microsoft Azure, AWS, and Google Cloud to public and regulatory pressure, which have included suspending sales of FRT tools and revisiting internal ethical guidelines \cite{Wen&Holweg2024}.

\section{Methodology}

\begin{figure}[h!]
    \centering
    \includegraphics[width=0.5\textwidth]{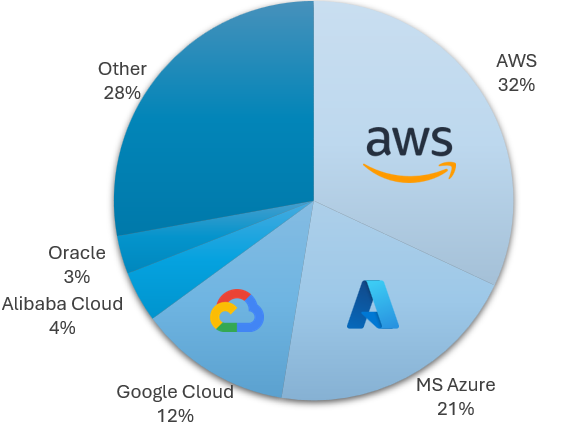}
    \caption{Cloud providers market share}
    \label{fig:market_share}
\end{figure}

This study conducts a comparative analysis of how three leading cloud service providers - Microsoft Azure, AWS, and Google Cloud- address the privacy principle in relation to facial recognition technologies. These cloud platforms were selected based on their significant market influence. According to Statista's report from Q3 2024 on the worldwide market share of the leading cloud service providers \cite{statista2025}, AWS, Microsoft Azure, and Google Cloud collectively dominate the cloud infrastructure market, which is represented in Figure \ref{fig:market_share}.

Data was collected from official company sources, including guidelines, technical documentation, policies, public announcements, as well as external media articles discussing the companies' approaches regarding FRT privacy, which are summarized in Table \ref{tab:sources_providers}. The selected timeframe, spanning from 2018 to 2024, captures key developments in privacy policies following increased regulatory scrutiny and public awareness of AI ethics.

\begin{table}[h!] 
  \caption{Distribution of sources by cloud provider}
  \label{tab:sources_providers}
  \begin{tabular}{p{3cm} p{4cm}}
    \toprule
    \bfseries Cloud Provider&  \bfseries Number of Sources\\
    \midrule
    AWS&    16\\
 Microsoft& 15\\
    Google&     10\\
    \bottomrule
\end{tabular}
\end{table}

The practical analysis is structured into the following parts:

\begin{itemize}
    \item Shared Responsibility Model adopted by the cloud providers, which defines the division of privacy and security responsibilities between providers and customers. 
    \item FRT-specific privacy measures, which analysis and compares each provider's specific approach to FRT, including restrictions and service capabilities, data protection tools, and additional FRT-specific guidelines provided by cloud providers. 
\end{itemize}

\section{Discussions}

\subsection{Data Protection in the Cloud: Shared Responsibility Model}

The three cloud providers (Microsoft Azure, AWS, and Google)
implement the shared responsibility model\footnote{In this regards, it is important to clarify the following terms:
\begin{itemize}
    \item \textbf{The customer} is the entity purchasing cloud services (whether a company or an individual) \cite{nist800-146}. In the context of data privacy, the customer acts as the \textbf{data controller}, i.e., determines the purposes and means of processing personal data \cite{aws_gdpr}. 
    \item \textbf{Cloud providers} act as \textbf{data processors}, processing data on behalf of data controllers \cite{aws_gdpr}. 
 \end{itemize}} for cloud security, where customers retain ownership of the data, while the cloud provider acts as the data processor, responsible for managing the infrastructure and services that support the data \cite{MicrosoftSharedResponsibility, AWS2024DataProtection, google2023shared}. 

In this model, the cloud provider is responsible for the security of the cloud infrastructure, including physical security, network controls, and the underlying hardware and software. 

Customers are responsible for the following:
\begin{itemize}
    \item maintaining and securing their data, 
    \item access control and identity management,
    \item configuring security settings for their applications and services,
    \item ensuring that their use of services is in accordance with relevant legal frameworks, such as GDPR and CCPA \cite{MicrosoftSharedResponsibility, AWS2024DataProtection, google2023shared}.
\end{itemize}

To support this, cloud providers offer a range of tools and services that will be examined in dedicated subsections. 

In their turn, cloud providers take measures to preserve the confidentiality of customer data. Some key measures Microsoft takes to protect the confidentiality of customers' data include the following: 
\begin{itemize}
    \item All three providers do not use customer data for advertising or marketing without explicit consent, but Google's policy explicitly mentions not selling personal information. 
    \item All three providers do not disclose customer content unless legally required, with attempts to notify or redirect requests to the customer unless legally prohibited.
    \item All three providers comply with international standard for data protection, including ISO 27018, a global standard for protecting Personally Identifiable Information in public cloud environments \cite{microsoft_assurance_privacy, aws_data_privacy_faq, google_privacy_policy}.
\end{itemize}

\subsection{Microsoft Azure's Privacy Approach to Facial Recognition}

\subsubsection{Azure AI Face Service}
The Microsoft Azure AI Face Service, which was introduced in 2015 as part of Microsoft's Cognitive Services, provides advanced facial recognition capabilities that allow users to detect, recognize, and analyse human faces within images. Its APIs, summarized in table \ref{tab:azure_face_apis}, provide diverse functionalities \cite{Microsoft2024Face_API}:

\begin{table}[h!] 
  \caption{Microsoft Azure Face APIs}
  \label{tab:azure_face_apis}
  \begin{tabular}{p{3cm} p{4cm}}
    \toprule
    \bfseries API&  \bfseries Description\\
    \midrule
    Verification&    Compares two faces to determine if they belong to the same person.\\
 Identification& Matches a detected face against a database to identify the individual\\
    Detection&     Detects faces in images and provides basic attributes (location, landmarks, etc.).\\
 Find Similar&Finds similar faces from a database by comparing a given face to others.\\
 Group&Groups similar faces based on visual similarity without pre-defined labels.\\
 Face Liveness&Detects whether a face is from a live person or a spoof attempt\\
 \bottomrule
\end{tabular}
\end{table}

However, not all features remain publicly available. In 2022, Microsoft announced that access to advanced facial recognition functionalities—Verification, Identification, Find Similar, Group, and Face Liveness—requires customers to submit their use cases for approval before use \cite{Bird2024, MicrosoftLimitedCapabilities}. 

Among the publicly available capabilities in Face Detection are attributes like blur detection, head pose estimation, landmarks identification, and glasses detection. However, Microsoft fully retired emotion and gender detection due to concerns over privacy risks, potential misuse, and the lack of scientific consensus on defining emotions \cite{MicrosoftLimitedCapabilities}. As Sarah Bird, Microsoft's Responsible AI Lead, highlighted, emotion classification posed challenges due to the subjective nature of emotions, the inability to generalize facial expressions across regions and demographics, and the risk of stereotyping or bias \cite{Bird2024}.

Additionally, features like age, smile, facial hair, and makeup detection have been restricted and now require customers to demonstrate a responsible use case for access \cite{MicrosoftLimitedCapabilities}. 

Moreover, the restricted use approval process for Azure AI Face requires periodic renewal, which allows Microsoft to continuously monitor and evaluate the responsible use of facial recognition service by customers \cite{MSLimitedAccessFaceApi}.

\subsubsection{Data Protection Tools}
Customers are recommended to use the following tools to protect their data in general and in particular in the context of AI-driven facial recognition and detection services, as outlined in Table \ref{tab:protection_tools_azure}.

\begin{table}[h!] 
  \caption{Microsoft Azure: Data Protection Tools}
  \label{tab:protection_tools_azure}
  \centering
  \renewcommand{\arraystretch}{1.2} 
  \begin{tabular}{p{3cm} p{4cm} }
    \toprule
    \bfseries Type of FRT&  \bfseries Microsoft Azure\\
    \midrule
    \bfseries Encryption services&    Encryption at rest and in transit, customer-managed keys \cite{Microsoft_Face_Privacy}\\
 \bfseries Network security& Private endpoint access, firewall rules for selected networks and IPs \cite{Microsoft_Face_Privacy}\\
    \bfseries Authentication and credential management&     Microsoft Entra ID \cite{Microsoft_Face_Privacy}\\
 \bfseries Access control& Azure Role Based Access Control (RBAC) for restricted operations \cite{Microsoft_Face_Privacy}\\
 \bfseries Data location control& Azure Regions for the  service and data storage and processing \cite{microsoft_data_residency}\\
 \bfseries Data discovery and protection& Azure Purview for data governance, discovery, and protection of sensitive biometric data \cite{microsoft_purview}.\\

\bottomrule
\end{tabular}

\end{table}

\subsubsection{Privacy Guidelines}
In addition with providing tools for safeguarding the data, Microsoft provides customers with comprehensive use guidelines and code of conduct to ensure that its Face API is used responsibly.  
\begin{itemize}
    \item "Guidance for integration and responsible use of Face" \cite{microsoft_face_guidance}
    \item "Code of conduct for Azure AI Vision Face API" \cite{microsoft_face_code_of_conduct}
    \item "Data and privacy for Face" \cite{Microsoft_Face_Privacy}
    \item "Best Practices for Adding Users to a Face Service" \cite{microsoft_computer_vision_enrollment}
\end{itemize}

Key privacy-related activities include:

\begin{itemize}
    \item Customers must collect biometric data only with explicit consent of the data subject \cite{microsoft_face_guidance}.
    \item The API must not be used to infer sensitive attributes like emotional state or gender \cite{microsoft_face_code_of_conduct}.
    \item Human oversight is required in decision-making processes, particularly in cases where the outcomes affect an individual's legal rights or well-being  \cite{microsoft_face_guidance}.
    \item The customers are suggested to implement strict data retention plans (e.g. deleting raw data once insights are derived) \cite{microsoft_face_guidance}.
    \item Customers must secure any data collected. This includes preventing unauthorized access, de-identification, and encryption \cite{microsoft_face_guidance}.
    \item The service cannot be used for real-time surveillance, persistent tracking, or identification without valid consent, especially in law enforcement applications \cite{microsoft_face_code_of_conduct}.
    \item Use of the service for individuals under the age of consent or in ways that could exploit or manipulate minors is prohibited \cite{microsoft_face_code_of_conduct}.
\end{itemize}

Additionally, in June 2020, following widespread protests against police practices in the United States, Microsoft announced a ban on the use of Azure AI Face technology by or for U.S. police departments \cite{Microsoft2024Transparency}. 

Microsoft facilitates compliance by offering customers detailed information on national, regional, and industry-specific regulations governing data collection and usage. For instance, the \textit{Microsoft Compliance Offerings} portal provides comprehensive guidance on 32 regional, 16 global, 14 US government, and 38 industry-specific regulations, assisting customers in meeting their compliance obligations \cite{microsoft_compliance_offerings}. Many of of these standards and regulations are explicitly focused on privacy: ISO 27018, HIPAA/ HITECH, PCI-DSS, GDPR, CCPA, Canadian Privacy Laws, etc. 

\subsection{AWS' Privacy Approach to Facial Recognition}

\subsubsection{Amazon Rekognition}

Launched by AWS in 2016, Amazon Rekognition provides advanced facial recognition capabilities, enabling users to detect, analyze and compare faces for the variety of applications. Key features are resported in Table \ref{tab:Amazon_Recognition_APIs}: 

\begin{table}[h!] 
  \caption{Amazon Rekognition APIs}
  \label{tab:Amazon_Recognition_APIs}
  \begin{tabular}{p{3cm} p{4cm}}
    \toprule
    \bfseries API&  \bfseries Description\\
    \midrule
    Face detection&    Identifies faces within images and videos \cite{AWS_Face_types}.\\
 Face comparison& Compares two faces to determine similarity \cite{AWS_Face_types}.\\
    Facial analysis&     Detects facial attributes such as age range, gender, emotions, and facial landmarks \cite{aws_face_attributes}.\\
 Face search&Searches for faces in a collection to find matches \cite{aws_collections}.\\
 Celebrity recognition&Identifies celebrities in images or videos \cite{aws_celebrities}.\\
\bottomrule
\end{tabular}
\end{table}

However, unlike Azure AI Face Service, Amazon Rekognition does not restrict access to specific features, and any AWS account holder can utilize APIs. 

In June 2020, also Amazon announced a one-year moratorium on police use of \textit{Face Comparison} and \textit{Face Search} features of Rekognition, responding to concerns about potential misuse and privacy implications \cite{aws_face_matching, Hao2020}. 

This decision followed not only protests against actions of the US police force, but also scrutiny from civil liberties organizations, such as the American Civil Liberties Union, which reported in 2018 that Rekognition had incorrectly matched 28 members of Congress with mugshots, disproportionally affecting people of colour \cite{aclu2020}.

In June 2021, Amazon extended the moratorium indefinitely \cite{aclu_extended2021}.

\subsubsection{Data Protection Tools}

AWS recommends using the following tools to protect sensitive data, including Amazon Rekognition:

\begin{table}[h!] 
  \caption{AWS:Data Protection Tools}
  \label{tab:protection_tools_AWS}
  \centering
  \renewcommand{\arraystretch}{1.2} 
  \begin{tabular}{p{3cm} p{4cm}}
    \toprule
    \bfseries Tool& \bfseries AWS\\
    \midrule
    \bfseries Encryption services&Encryption at rest and in transit, customer-managed keys \cite{AWS2024DataProtection}\\
 \bfseries Network security&Amazon VPC endpoints for isolated and secure network connections \cite{aws_rekognition_vpc}\\
    \bfseries Authentication and credential management&AWS Identity and Access Management (IAM) \cite{aws_rekognition_IAM}\\
 \bfseries Access control& AWS IAM Roles and Policies \cite{aws_rekognition_IAM}\\
 \bfseries Data location control& AWS Regions for the  service and data storage and processing \cite{aws_rekognition_regions}\\
 \bfseries Data discovery and protection& Amazon Macie to monitor, identify, and secure sensitive biometric data stored in Amazon S3 \cite{AWS2024DataProtection}\\

\bottomrule
\end{tabular}

\end{table}

\subsubsection{Privacy Guidelines}

In addition to the recommended tools to protect the data, AWS advises customers to implement the following measures to meet privacy requirements:
\begin{itemize}
    \item Ensuring that users and applications are granted only the minimum permissions to perform their task.
    \item Avoiding storing sensitive information in tags and free-form fields \cite{AWS2024DataProtection}
\end{itemize}

AWS offers a Compliance Center portal covering 69 countries, reflecting the geographic reach of its services \cite{aws_compliance_center}. Through this portal, customers can find information on laws relevant to privacy and data protection within a selected country.  

\subsection{Google Cloud's Privacy Approach to Facial Recognition}

\subsubsection{Cloud Vision API: Face Detection and Celebrity Recognition}

Unlike its major competitors, Google has chosen not to offer general-purpose facial recognition services. Instead, they provide face detection, which also includes emotion recognition capabilities, through their Cloud Vision API and ML Kit. Specifically, Face Detection response includes bounding boxes for faces, landmarks detected, and confidence ratings for the following expressions: joy, sorrow, anger, and surprise \cite{google_vision_faces}. 

Another Face recognition related feature offered by Google Cloud is \textit{Celebrity Recognition}. The Cloud Vision API identifies the celebrity by comparing the face with an indexed gallery of celebrities. However, this feature is deprecated and will not be available after September 16, 2025 \cite{google_vision_celebrity_recognition}.

\subsubsection{Data Protection Tools}

Although Goole Cloud does not provide the service-specific responsible use recommendations, the following tools are recommended to secure sensitive data across its services in general:

\begin{table}[h!] 
  \caption{Google Cloud: Data Protection Tools}
  \label{tab:protection_tools_google}
  \centering
  \renewcommand{\arraystretch}{1.2} 
  \begin{tabular}{p{3cm} p{4cm}}
    \toprule
    \bfseries Type of FRT& \bfseries GCP\\
    \midrule
    \bfseries Encryption services&Encryption at rest and in transit, customer-managed keys \cite{googlecloudsecurity}\\
 \bfseries Network security& VPC Service Control for API-level security \cite{google_vpc_service_controls}\\
    \bfseries Authentication and credential management& Google Cloud Identity and Access Management (IAM) \cite{google_cloud_iam}\\
 \bfseries Access control&Google Cloud IAM Roles and Policies \cite{google_cloud_iam}\\
 \bfseries Data location control&Google Vision API uses global endpoints, but customers can choose Google Cloud Regions for storing and processing the data \cite{google_geography}\\
 \bfseries Data discovery and protection&Google Cloud Sensitive Data Protection for discovering, classifying, and securing sensitive data \cite{google_cloud_sensitive}\\

\bottomrule
\end{tabular}

\end{table}

\subsubsection{Privacy Guidelines}

While Google provides general guidelines on data privacy and security broadly across its services, specific responsible use guidelines for Cloud Vision API are not explicitly detailed. 

Google Cloud offers a Compliance Center that offers detailed information on various regulations. Users can filter the data by country, region, industry, and focus area. When filtering by the 'Privacy' focus area, 37 relevant regulations are identified \cite{google_cloud_compliance_center}. 

\subsection{Comparison of Approaches}

Beyond traditional facial recognition functions such as verification, identification, categorization, and emotion detection found in the systematic literature review, additional features can be found among the ones offered by the reviewed cloud providers. For example, Microsoft offers "Find Similar" and "Face Liveness" features, AWS provides age range and gender detection, and both AWS and Google Cloud offer celebrity recognition capabilities. 

There are significant differences in the features available and the level of access:
\begin{itemize}
    \item \textbf{Microsoft Azure} limits access to its facial recognition features. Customers must submit use cases for approval, and these cases are periodically revalidated to ensure responsible use. Microsoft has also retired emotion and gender recognition features from its face detection service.  This approach is more restrictive, prioritizing privacy by limiting the scope of the use of technology. 
    \item \textbf{AWS}, by contrast, offers broad availability of its facial recognition features. Any user with an AWS account can access and use these tools without stringent approval processes. While this provides flexibility, it also increases the privacy risks, as it could lead to unauthorized use and surveillance without consent.
    \item \textbf{Google Cloud} has chosen a middle-ground approach. While it does not offer general-purpose facial recognition services, it does provide emotion recognition. Google has also opted to retire its celebrity recognition feature by 2025. By restricting facial recognition use and focusing on face detection, Google mitigates some privacy concerns, though its emotion recognition feature still raises potential issues, particularly if used in unauthorized contexts (for example, in the context of a job interview or customer interactions and without an explicit consent from a data subject). 
\end{itemize}

The three cloud providers—Microsoft Azure, AWS, and Google Cloud—offer compliance centres that provide detailed information on various regulations.  

In addition to compliance resources, providers offer a range of tools designed to protect customer data. These include data encryption, access controls, storage location, which were mentioned in systematic literature review. However, the specific types of encryption employed are not specified in the available documentation. Further tools, not covered in the review, include network restrictions and specific tools for monitoring and protecting sensitive and biometric data. 

Microsoft stands out by offering clear privacy and responsible use guidelines specifically dedicated to its face recognition services. These guidelines help customers adhere to best practices for privacy and security when implementing facial recognition technology.

\section{Conclusion}

The paper studied how privacy is addressed in cloud-based facial recognition services by major cloud providers Microsoft Azure, AWS, and Google Cloud - within the framework of Responsible AI. 

The systematic literature review reinforced that privacy is one of the key principles of Responsible AI and underscores a commitment to protecting sensitive data specifically in the cloud environment. 

The study revealed significant variations in privacy approaches, where Microsoft applies restrictive and case-revalidation usage approach, AWS offers broad access by shifting greater responsibilities on the customer and increasing privacy risks, and Google adopts a cautious approach by focusing on face detection with some emotion recognition functionalities. 

Key insights demonstrate that, in addition to usage restriction and not selling the service, privacy measures are implemented through: 
\begin{itemize}
    \item \textbf{technical tools}, such as data encryption, access control, storage management, networking restrictions, and sensitive data protection services are universally employed.
    \item \textbf{regulatory awareness}: informing customers on relevant regulations through compliance centres.
    \item \textbf{privacy guidelines}: the three cloud providers offer general privacy guidelines. Microsoft additionally provides specific guidelines dedicated to on responsible use of facial recognition services. 
\end{itemize}

Furthermore, under the shared responsibility model, ultimate responsibility for ensuring privacy rests with customers and their choices in implementing these tools, guidelines and relevant regulations. 

The findings of this study hold valuable implications for the broader landscape of Responsible AI. In the immediate term, they provide meaningful insights into varied approaches adopted by cloud providers, enabling customers to make informed decisions when developing facial recognition systems. These insights also brings customer's attention to the areas that they are responsible for and concrete means available to them to address privacy.  Long-term impacts may include the need for more standardized privacy practices across cloud providers to reduce variability and ambiguity for customers.  

While this study provides valuable insights, questions remain due to the specified focus of the study. The following future research directions can expand the knowledge base and provide valuable insights for industry practices:
\begin{itemize}
    \item A comprehensive study of the key Responsible AI principles (e.g. privacy, transparency, fairness) applied collectively to cloud-based facial recognition services could offer a more holistic view. 
    \item Future research could also explore the interplay between privacy and transparency or privacy and fairness, examining how these principles complement or conflict with each other in practical applications.
    \item Future research can also explore privacy implication from the customer's perspective, focusing on a case study of a specific cloud-based FRT application(s).This could provide deeper insights into how privacy measures are applied on the client's side, including tools utilized, regulatory compliance, implementation challenges, and overall effectiveness of privacy safeguards. 
\end{itemize}


\bibliographystyle{ACM-Reference-Format}
\bibliography{sample-base}


\begin{thebibliography}{70}


\ifx \showCODEN    \undefined \def \showCODEN     #1{\unskip}     \fi
\ifx \showDOI      \undefined \def \showDOI       #1{#1}\fi
\ifx \showISBNx    \undefined \def \showISBNx     #1{\unskip}     \fi
\ifx \showISBNxiii \undefined \def \showISBNxiii  #1{\unskip}     \fi
\ifx \showISSN     \undefined \def \showISSN      #1{\unskip}     \fi
\ifx \showLCCN     \undefined \def \showLCCN      #1{\unskip}     \fi
\ifx \shownote     \undefined \def \shownote      #1{#1}          \fi
\ifx \showarticletitle \undefined \def \showarticletitle #1{#1}   \fi
\ifx \showURL      \undefined \def \showURL       {\relax}        \fi
\providecommand\bibfield[2]{#2}
\providecommand\bibinfo[2]{#2}
\providecommand\natexlab[1]{#1}
\providecommand\showeprint[2][]{arXiv:#2}

\bibitem[Bird(2022)]%
        {Bird2024}
\bibfield{author}{\bibinfo{person}{Sarah Bird}.} \bibinfo{year}{2022}\natexlab{}.
\newblock \bibinfo{booktitle}{\emph{Responsible AI Investments and Safeguards for Facial Recognition}}.
\newblock
\urldef\tempurl%
\url{https://azure.microsoft.com/en-us/blog/responsible-ai-investments-and-safeguards-for-facial-recognition/}
\showURL{%
\tempurl}


\bibitem[Butt et~al\mbox{.}(2023)]%
        {Butt2023}
\bibfield{author}{\bibinfo{person}{Muhammad~A. Butt}, \bibinfo{person}{Adnan Qayyum}, \bibinfo{person}{Hassan Ali}, \bibinfo{person}{Ala Al-Fuqaha}, {and} \bibinfo{person}{Junaid Qadir}.} \bibinfo{year}{2023}\natexlab{}.
\newblock \showarticletitle{Towards secure private and trustworthy human-centric embedded machine learning: An emotion-aware facial recognition case study}.
\newblock \bibinfo{journal}{\emph{Computers \& security}}  \bibinfo{volume}{125} (\bibinfo{year}{2023}), \bibinfo{pages}{103058}.
\newblock
\showISBNx{0167-4048}


\bibitem[Buyukcavus et~al\mbox{.}(2023)]%
        {Buyukcavus2023}
\bibfield{author}{\bibinfo{person}{Muhammed~H. Buyukcavus}, \bibinfo{person}{Filiz Aydogan~Akgun}, \bibinfo{person}{Serdar Solak}, \bibinfo{person}{Mustafa H.~B. Ucar}, \bibinfo{person}{Yavuz Fındık}, {and} \bibinfo{person}{Timucin Baykul}.} \bibinfo{year}{2023}\natexlab{}.
\newblock \showarticletitle{Facial recognition by cloud-based APIs following surgically assisted rapid maxillary expansion}.
\newblock \bibinfo{journal}{\emph{Journal of orofacial orthopedics}} (\bibinfo{year}{2023}).
\newblock
\showISBNx{1434-5293}


\bibitem[Cheng et~al\mbox{.}(2021)]%
        {Cheng}
\bibfield{author}{\bibinfo{person}{Lu Cheng}, \bibinfo{person}{Kush~R. Varshney}, {and} \bibinfo{person}{Huan Liu}.} \bibinfo{year}{2021}\natexlab{}.
\newblock \showarticletitle{Socially Responsible {AI} Algorithms: Issues, Purposes, and Challenges}.
\newblock \bibinfo{journal}{\emph{CoRR}}  \bibinfo{volume}{abs/2101.02032} (\bibinfo{year}{2021}).
\newblock
\showeprint[arXiv]{2101.02032}
\urldef\tempurl%
\url{https://arxiv.org/abs/2101.02032}
\showURL{%
\tempurl}


\bibitem[Choung et~al\mbox{.}(2024)]%
        {Hyesun2024}
\bibfield{author}{\bibinfo{person}{Hyesun Choung}, \bibinfo{person}{Prabu David}, {and} \bibinfo{person}{Tsai-Wei Ling}.} \bibinfo{year}{2024}\natexlab{}.
\newblock \showarticletitle{Acceptance of AI-powered facial recognition technology in surveillance scenarios: Role of trust, security, and privacy perceptions}.
\newblock \bibinfo{journal}{\emph{Technology in society}}  \bibinfo{volume}{79} (\bibinfo{year}{2024}), \bibinfo{pages}{102721}.
\newblock
\showISBNx{0160-791X}


\bibitem[Clarke(2019)]%
        {CLARKE2019}
\bibfield{author}{\bibinfo{person}{Roger Clarke}.} \bibinfo{year}{2019}\natexlab{}.
\newblock \showarticletitle{Principles and business processes for responsible AI}.
\newblock \bibinfo{journal}{\emph{Computer Law \& Security Review}} \bibinfo{volume}{35}, \bibinfo{number}{4} (\bibinfo{year}{2019}), \bibinfo{pages}{410--422}.
\newblock
\showISSN{0267-3649}
\urldef\tempurl%
\url{https://doi.org/10.1016/j.clsr.2019.04.007}
\showDOI{\tempurl}


\bibitem[Cloud(2023)]%
        {google2023shared}
\bibfield{author}{\bibinfo{person}{Google Cloud}.} \bibinfo{year}{2023}\natexlab{}.
\newblock \bibinfo{title}{Shared responsibilities and shared fate on Google Cloud}.
\newblock
\newblock
\urldef\tempurl%
\url{https://cloud.google.com/architecture/framework/security/shared-responsibility-shared-fate}
\showURL{%
\tempurl}


\bibitem[Cloud(2024a)]%
        {google_vision_celebrity_recognition}
\bibfield{author}{\bibinfo{person}{Google Cloud}.} \bibinfo{year}{2024}\natexlab{a}.
\newblock \bibinfo{title}{Celebrity Recognition with the Cloud Vision API}.
\newblock
\newblock
\urldef\tempurl%
\url{https://cloud.google.com/vision/docs/celebrity-recognition}
\showURL{%
\tempurl}


\bibitem[Cloud(2024b)]%
        {google_vision_faces}
\bibfield{author}{\bibinfo{person}{Google Cloud}.} \bibinfo{year}{2024}\natexlab{b}.
\newblock \bibinfo{title}{Detecting Faces with the Cloud Vision API}.
\newblock
\newblock
\urldef\tempurl%
\url{https://cloud.google.com/vision/docs/detecting-faces}
\showURL{%
\tempurl}


\bibitem[Cloud(2024c)]%
        {google_cloud_compliance_center}
\bibfield{author}{\bibinfo{person}{Google Cloud}.} \bibinfo{year}{2024}\natexlab{c}.
\newblock \bibinfo{title}{Google Cloud Compliance Center}.
\newblock
\newblock
\urldef\tempurl%
\url{https://cloud.google.com/security/compliance}
\showURL{%
\tempurl}


\bibitem[Cloud(2024d)]%
        {google_geography}
\bibfield{author}{\bibinfo{person}{Google Cloud}.} \bibinfo{year}{2024}\natexlab{d}.
\newblock \bibinfo{title}{Google Cloud Geography and Regions Overview}.
\newblock
\newblock
\urldef\tempurl%
\url{https://cloud.google.com/docs/geography-and-regions}
\showURL{%
\tempurl}


\bibitem[Cloud(2024e)]%
        {google_cloud_iam}
\bibfield{author}{\bibinfo{person}{Google Cloud}.} \bibinfo{year}{2024}\natexlab{e}.
\newblock \bibinfo{title}{Google Cloud IAM Overview}.
\newblock
\newblock
\urldef\tempurl%
\url{https://cloud.google.com/iam/docs/overview}
\showURL{%
\tempurl}


\bibitem[Cloud(2024f)]%
        {googlecloudsecurity}
\bibfield{author}{\bibinfo{person}{Google Cloud}.} \bibinfo{year}{2024}\natexlab{f}.
\newblock \bibinfo{title}{Google Cloud Security Overview}.
\newblock
\newblock
\urldef\tempurl%
\url{https://cloud.google.com/docs/security/overview/whitepaper}
\showURL{%
\tempurl}


\bibitem[Cloud(2024g)]%
        {google_cloud_sensitive}
\bibfield{author}{\bibinfo{person}{Google Cloud}.} \bibinfo{year}{2024}\natexlab{g}.
\newblock \bibinfo{title}{Google Cloud Sensitive Data Protection Overview}.
\newblock
\newblock
\urldef\tempurl%
\url{https://cloud.google.com/sensitive-data-protection/docs/sensitive-data-protection-overview}
\showURL{%
\tempurl}


\bibitem[Cloud(2024h)]%
        {google_vpc_service_controls}
\bibfield{author}{\bibinfo{person}{Google Cloud}.} \bibinfo{year}{2024}\natexlab{h}.
\newblock \bibinfo{title}{Google Cloud VPC Service Controls Overview}.
\newblock
\newblock
\urldef\tempurl%
\url{https://cloud.google.com/vpc-service-controls/docs/overview}
\showURL{%
\tempurl}


\bibitem[Commission(2024)]%
        {eu_ai_act}
\bibfield{author}{\bibinfo{person}{European Commission}.} \bibinfo{year}{2024}\natexlab{}.
\newblock \bibinfo{title}{Regulation of the European Parliament and of the Council Laying Down Harmonised Rules on Artificial Intelligence (Artificial Intelligence Act)}.
\newblock
\newblock
\urldef\tempurl%
\url{https://eur-lex.europa.eu/legal-content/EN/TXT/?uri=CELEX%3A52021PC0206}
\showURL{%
\tempurl}


\bibitem[Dignum(2019)]%
        {dignum2019}
\bibfield{author}{\bibinfo{person}{Virginia Dignum}.} \bibinfo{year}{2019}\natexlab{}.
\newblock \bibinfo{booktitle}{\emph{Responsible Artificial Intelligence: How to Develop and Use AI in a Responsible Way} (\bibinfo{edition}{1st 2019.} ed.)}.
\newblock \bibinfo{publisher}{Springer}, \bibinfo{address}{Cham}.
\newblock
\showISBNx{9783030303716;3030303713;}


\bibitem[Domingo-Ferrer et~al\mbox{.}(2019)]%
        {DOMINGOFERRER201938}
\bibfield{author}{\bibinfo{person}{Josep Domingo-Ferrer}, \bibinfo{person}{Oriol Farràs}, \bibinfo{person}{Jordi Ribes-González}, {and} \bibinfo{person}{David Sánchez}.} \bibinfo{year}{2019}\natexlab{}.
\newblock \showarticletitle{Privacy-preserving cloud computing on sensitive data: A survey of methods, products and challenges}.
\newblock \bibinfo{journal}{\emph{Computer Communications}}  \bibinfo{volume}{140-141} (\bibinfo{year}{2019}), \bibinfo{pages}{38--60}.
\newblock
\showISSN{0140-3664}
\urldef\tempurl%
\url{https://doi.org/10.1016/j.comcom.2019.04.011}
\showDOI{\tempurl}


\bibitem[Díaz-Rodríguez et~al\mbox{.}(2023)]%
        {DIAZRODRIGUEZ}
\bibfield{author}{\bibinfo{person}{Natalia Díaz-Rodríguez}, \bibinfo{person}{Javier {Del Ser}}, \bibinfo{person}{Mark Coeckelbergh}, \bibinfo{person}{Marcos {López de Prado}}, \bibinfo{person}{Enrique Herrera-Viedma}, {and} \bibinfo{person}{Francisco Herrera}.} \bibinfo{year}{2023}\natexlab{}.
\newblock \showarticletitle{Connecting the dots in trustworthy Artificial Intelligence: From AI principles, ethics, and key requirements to responsible AI systems and regulation}.
\newblock \bibinfo{journal}{\emph{Information Fusion}}  \bibinfo{volume}{99} (\bibinfo{year}{2023}), \bibinfo{pages}{101896}.
\newblock
\showISSN{1566-2535}
\urldef\tempurl%
\url{https://doi.org/10.1016/j.inffus.2023.101896}
\showDOI{\tempurl}


\bibitem[Ferdowsi et~al\mbox{.}(2024)]%
        {FERDOWSI2024}
\bibfield{author}{\bibinfo{person}{Mahbuba Ferdowsi}, \bibinfo{person}{Md~Mahmudul Hasan}, {and} \bibinfo{person}{Wafa Habib}.} \bibinfo{year}{2024}\natexlab{}.
\newblock \showarticletitle{Responsible AI for cardiovascular disease detection: Towards a privacy-preserving and interpretable model}.
\newblock \bibinfo{journal}{\emph{Computer Methods and Programs in Biomedicine}}  \bibinfo{volume}{254} (\bibinfo{year}{2024}), \bibinfo{pages}{108289}.
\newblock
\showISSN{0169-2607}
\urldef\tempurl%
\url{https://doi.org/10.1016/j.cmpb.2024.108289}
\showDOI{\tempurl}


\bibitem[Fjeld et~al\mbox{.}(2020)]%
        {fjeld2020principled}
\bibfield{author}{\bibinfo{person}{Jessica Fjeld}, \bibinfo{person}{Nele Achten}, \bibinfo{person}{Hannah Hilligoss}, \bibinfo{person}{Adam Nagy}, {and} \bibinfo{person}{Madhulika Srikumar}.} \bibinfo{year}{2020}\natexlab{}.
\newblock \bibinfo{booktitle}{\emph{Principled Artificial Intelligence: Mapping Consensus in Ethical and Rights-Based Approaches to Principles for AI}}.
\newblock \bibinfo{type}{White Paper}. \bibinfo{institution}{Berkman Klein Center for Internet \& Society, Harvard Law School}.
\newblock


\bibitem[Goellner et~al\mbox{.}(2024)]%
        {goellner2024responsible}
\bibfield{author}{\bibinfo{person}{Sabrina Goellner}, \bibinfo{person}{Marina Tropmann-Frick}, {and} \bibinfo{person}{Bostjan Brumen}.} \bibinfo{year}{2024}\natexlab{}.
\newblock \bibinfo{title}{Responsible Artificial Intelligence: A Structured Literature Review}.
\newblock
\newblock
\showeprint[arxiv]{2403.06910}~[cs.AI]


\bibitem[Google(2024a)]%
        {google_ai_principles}
\bibfield{author}{\bibinfo{person}{Google}.} \bibinfo{year}{2024}\natexlab{a}.
\newblock \bibinfo{booktitle}{\emph{AI Principles}}.
\newblock
\urldef\tempurl%
\url{https://ai.google/responsibility/principles/}
\showURL{%
\tempurl}


\bibitem[Google(2024b)]%
        {google_privacy_policy}
\bibfield{author}{\bibinfo{person}{Google}.} \bibinfo{year}{2024}\natexlab{b}.
\newblock \bibinfo{title}{Google Privacy Policy}.
\newblock
\newblock
\urldef\tempurl%
\url{https://policies.google.com/privacy?hl=en-US}
\showURL{%
\tempurl}


\bibitem[Hao(2020)]%
        {Hao2020}
\bibfield{author}{\bibinfo{person}{Karen Hao}.} \bibinfo{year}{2020}\natexlab{}.
\newblock \showarticletitle{Amazon Stopped Selling Police Face Recognition}.
\newblock \bibinfo{journal}{\emph{MIT Technology Review}} (\bibinfo{year}{2020}).
\newblock
\urldef\tempurl%
\url{https://www.technologyreview.com/2020/06/12/1003482/amazon-stopped-selling-police-face-recognition-fight}
\showURL{%
\tempurl}


\bibitem[Hossain and Muhammad(2019)]%
        {HossainAnaMuhammad}
\bibfield{author}{\bibinfo{person}{M.~S. Hossain} {and} \bibinfo{person}{Ghulam Muhammad}.} \bibinfo{year}{2019}\natexlab{}.
\newblock \showarticletitle{Emotion recognition using secure edge and cloud computing}.
\newblock \bibinfo{journal}{\emph{Information sciences}}  \bibinfo{volume}{504} (\bibinfo{year}{2019}), \bibinfo{pages}{589--601}.
\newblock
\showISBNx{0020-0255}


\bibitem[IBM(2024)]%
        {ibm_responsible_ai}
\bibfield{author}{\bibinfo{person}{IBM}.} \bibinfo{year}{2024}\natexlab{}.
\newblock \bibinfo{booktitle}{\emph{Responsible AI}}.
\newblock
\urldef\tempurl%
\url{https://www.ibm.com/topics/responsible-ai}
\showURL{%
\tempurl}


\bibitem[Liu et~al\mbox{.}(2023)]%
        {Liu2023}
\bibfield{author}{\bibinfo{person}{Rui Liu}, \bibinfo{person}{Suraksha Gupta}, {and} \bibinfo{person}{Parth Patel}.} \bibinfo{year}{2023}\natexlab{}.
\newblock \showarticletitle{The Application of the Principles of Responsible AI on Social Media Marketing for Digital Health}.
\newblock \bibinfo{journal}{\emph{Information Systems Frontiers}} \bibinfo{volume}{25}, \bibinfo{number}{6} (\bibinfo{date}{December} \bibinfo{year}{2023}), \bibinfo{pages}{2275--2299}.
\newblock
\showISSN{1572-9419}
\urldef\tempurl%
\url{https://doi.org/10.1007/s10796-021-10191-z}
\showDOI{\tempurl}


\bibitem[Lu et~al\mbox{.}(2024)]%
        {Lu}
\bibfield{author}{\bibinfo{person}{Qinghua Lu}, \bibinfo{person}{Liming Zhu}, \bibinfo{person}{Xiwei Xu}, \bibinfo{person}{Jon Whittle}, \bibinfo{person}{Didar Zowghi}, {and} \bibinfo{person}{Aurelie Jacquet}.} \bibinfo{year}{2024}\natexlab{}.
\newblock \showarticletitle{Responsible AI Pattern Catalogue: A Collection of Best Practices for AI Governance and Engineering}.
\newblock \bibinfo{journal}{\emph{ACM Comput. Surv.}} \bibinfo{volume}{56}, \bibinfo{number}{7}, Article \bibinfo{articleno}{173} (\bibinfo{date}{April} \bibinfo{year}{2024}), \bibinfo{numpages}{35}~pages.
\newblock
\showISSN{0360-0300}
\urldef\tempurl%
\url{https://doi.org/10.1145/3626234}
\showDOI{\tempurl}


\bibitem[Matulionyte(2024)]%
        {Matulionyte2024}
\bibfield{author}{\bibinfo{person}{Rita Matulionyte}.} \bibinfo{year}{2024}\natexlab{}.
\newblock \showarticletitle{Increasing transparency around facial recognition technologies in law enforcement: towards a model framework}.
\newblock \bibinfo{journal}{\emph{Information \& communications technology law}} \bibinfo{volume}{33}, \bibinfo{number}{1} (\bibinfo{year}{2024}), \bibinfo{pages}{66--84}.
\newblock
\showISBNx{1360-0834}


\bibitem[Microsoft(2023a)]%
        {microsoft_data_residency}
\bibfield{author}{\bibinfo{person}{Microsoft}.} \bibinfo{year}{2023}\natexlab{a}.
\newblock \bibinfo{title}{Data Residency on Microsoft Azure}.
\newblock
\newblock
\urldef\tempurl%
\url{https://azure.microsoft.com/en-us/explore/global-infrastructure/data-residency}
\showURL{%
\tempurl}


\bibitem[Microsoft(2023b)]%
        {microsoft_purview}
\bibfield{author}{\bibinfo{person}{Microsoft}.} \bibinfo{year}{2023}\natexlab{b}.
\newblock \bibinfo{title}{Microsoft Purview Documentation}.
\newblock
\newblock
\urldef\tempurl%
\url{https://learn.microsoft.com/en-us/purview/purview}
\showURL{%
\tempurl}


\bibitem[Microsoft(2024a)]%
        {microsoft_ai_principles}
\bibfield{author}{\bibinfo{person}{Microsoft}.} \bibinfo{year}{2024}\natexlab{a}.
\newblock \bibinfo{booktitle}{\emph{AI Principles and Approach}}.
\newblock
\urldef\tempurl%
\url{https://www.microsoft.com/en-us/ai/principles-and-approach/?msockid=1c407d9617eb658331d96911169c64b0}
\showURL{%
\tempurl}


\bibitem[Microsoft(2024b)]%
        {microsoft_computer_vision_enrollment}
\bibfield{author}{\bibinfo{person}{Microsoft}.} \bibinfo{year}{2024}\natexlab{b}.
\newblock \bibinfo{title}{Best Practices to Adding Users to a Face Service}.
\newblock
\newblock
\urldef\tempurl%
\url{https://learn.microsoft.com/en-us/azure/ai-services/computer-vision/enrollment-overview}
\showURL{%
\tempurl}


\bibitem[Microsoft(2024c)]%
        {microsoft_face_code_of_conduct}
\bibfield{author}{\bibinfo{person}{Microsoft}.} \bibinfo{year}{2024}\natexlab{c}.
\newblock \bibinfo{title}{Code of Conduct for the Face service}.
\newblock
\newblock
\urldef\tempurl%
\url{https://learn.microsoft.com/en-us/legal/cognitive-services/face/code-of-conduct}
\showURL{%
\tempurl}


\bibitem[Microsoft(2024d)]%
        {Microsoft_Face_Privacy}
\bibfield{author}{\bibinfo{person}{Microsoft}.} \bibinfo{year}{2024}\natexlab{d}.
\newblock \bibinfo{booktitle}{\emph{Data and Privacy for Face}}.
\newblock
\urldef\tempurl%
\url{https://learn.microsoft.com/en-us/legal/cognitive-services/face/data-privacy-security}
\showURL{%
\tempurl}


\bibitem[Microsoft(2024e)]%
        {Microsoft2024Transparency}
\bibfield{author}{\bibinfo{person}{Microsoft}.} \bibinfo{year}{2024}\natexlab{e}.
\newblock \bibinfo{booktitle}{\emph{Face API Transparency Note}}.
\newblock
\urldef\tempurl%
\url{https://learn.microsoft.com/en-us/legal/cognitive-services/face/transparency-note?context=%2Fazure%2Fai-services%2Fcomputer-vision%2Fcontext%2Fcontext}
\showURL{%
\tempurl}


\bibitem[Microsoft(2024f)]%
        {MicrosoftLimitedCapabilities}
\bibfield{author}{\bibinfo{person}{Microsoft}.} \bibinfo{year}{2024}\natexlab{f}.
\newblock \bibinfo{booktitle}{\emph{Face Detection Concepts}}.
\newblock
\urldef\tempurl%
\url{https://learn.microsoft.com/en-us/azure/ai-services/computer-vision/concept-face-detection}
\showURL{%
\tempurl}


\bibitem[Microsoft(2024g)]%
        {microsoft_face_guidance}
\bibfield{author}{\bibinfo{person}{Microsoft}.} \bibinfo{year}{2024}\natexlab{g}.
\newblock \bibinfo{title}{Guidance for responsible integration of the Face service}.
\newblock
\newblock
\urldef\tempurl%
\url{https://learn.microsoft.com/en-us/legal/cognitive-services/face/guidance-integration-responsible-use}
\showURL{%
\tempurl}


\bibitem[Microsoft(2024h)]%
        {Microsoft2024Face_API}
\bibfield{author}{\bibinfo{person}{Microsoft}.} \bibinfo{year}{2024}\natexlab{h}.
\newblock \bibinfo{booktitle}{\emph{Identity API Reference}}.
\newblock
\urldef\tempurl%
\url{https://learn.microsoft.com/en-us/azure/ai-services/computer-vision/identity-api-reference}
\showURL{%
\tempurl}


\bibitem[Microsoft(2024i)]%
        {MSLimitedAccessFaceApi}
\bibfield{author}{\bibinfo{person}{Microsoft}.} \bibinfo{year}{2024}\natexlab{i}.
\newblock \bibinfo{booktitle}{\emph{Limited Access Identity}}.
\newblock
\urldef\tempurl%
\url{https://learn.microsoft.com/en-us/legal/cognitive-services/computer-vision/limited-access-identity?context=%2Fazure%2Fai-services%2Fcomputer-vision%2Fcontext%2Fcontext}
\showURL{%
\tempurl}


\bibitem[{Microsoft}(2024)]%
        {microsoft_compliance_offerings}
\bibfield{author}{\bibinfo{person}{{Microsoft}}.} \bibinfo{year}{2024}\natexlab{}.
\newblock \bibinfo{booktitle}{\emph{{Microsoft Compliance Offerings}}}.
\newblock
\urldef\tempurl%
\url{https://learn.microsoft.com/en-us/compliance/regulatory/offering-home}
\showURL{%
\tempurl}


\bibitem[Microsoft(2024a)]%
        {microsoft_assurance_privacy}
\bibfield{author}{\bibinfo{person}{Microsoft}.} \bibinfo{year}{2024}\natexlab{a}.
\newblock \bibinfo{title}{Microsoft Privacy Assurance}.
\newblock
\newblock
\urldef\tempurl%
\url{https://learn.microsoft.com/en-us/compliance/assurance/assurance-privacy}
\showURL{%
\tempurl}


\bibitem[Microsoft(2024b)]%
        {MicrosoftSharedResponsibility}
\bibfield{author}{\bibinfo{person}{Microsoft}.} \bibinfo{year}{2024}\natexlab{b}.
\newblock \bibinfo{title}{Shared Responsibility in the Cloud}.
\newblock
\newblock
\urldef\tempurl%
\url{https://learn.microsoft.com/en-us/azure/security/fundamentals/shared-responsibility}
\showURL{%
\tempurl}


\bibitem[{National Institute of Standards and Technology}(2012)]%
        {nist800-146}
\bibfield{author}{\bibinfo{person}{{National Institute of Standards and Technology}}.} \bibinfo{year}{2012}\natexlab{}.
\newblock \bibinfo{booktitle}{\emph{Cloud Computing Synopsis and Recommendations}}.
\newblock \bibinfo{type}{{T}echnical {R}eport} NIST SP 800-146. \bibinfo{institution}{U.S. Department of Commerce}.
\newblock
\urldef\tempurl%
\url{https://nvlpubs.nist.gov/nistpubs/Legacy/SP/nistspecialpublication800-146.pdf}
\showURL{%
\tempurl}


\bibitem[Nwafor(2023)]%
        {Nwafor2023}
\bibfield{author}{\bibinfo{person}{Ifeoma~E. Nwafor}.} \bibinfo{year}{2023}\natexlab{}.
\newblock \showarticletitle{Artificial Intelligence Facial Recognition Surveillance and the Breach of Privacy Rights: the ‘Clearview Ai’ and ‘Rite Aid’ Case Studies}.
\newblock \bibinfo{journal}{\emph{South African intellectual property law journal}} \bibinfo{volume}{11}, \bibinfo{number}{1} (\bibinfo{year}{2023}), \bibinfo{pages}{88}.
\newblock
\showISBNx{2309-4532}


\bibitem[Okon et~al\mbox{.}(2024)]%
        {Okon2024}
\bibfield{author}{\bibinfo{person}{Samuel~U. Okon}, \bibinfo{person}{Omobolaji~O. Olateju}, \bibinfo{person}{Olumide~S. Ogungbemi}, \bibinfo{person}{Sunday~A. Joseph}, \bibinfo{person}{Anthony~O. Olisa}, {and} \bibinfo{person}{Oluwaseun~O. Olaniyi}.} \bibinfo{year}{2024}\natexlab{}.
\newblock \showarticletitle{Incorporating Privacy by Design Principles in the Modification of AI Systems in Preventing Breaches across Multiple Environments, Including Public Cloud, Private Cloud, and On-prem}.
\newblock \bibinfo{journal}{\emph{Journal of Engineering Research and Reports}} \bibinfo{volume}{26}, \bibinfo{number}{9} (\bibinfo{year}{2024}), \bibinfo{pages}{136--158}.
\newblock
\showISBNx{2582-2926}


\bibitem[Patrick~Mikalef and Popovič(2022)]%
        {mikalef}
\bibfield{author}{\bibinfo{person}{Jenny Eriksson~Lundström Patrick~Mikalef, Kieran~Conboy} {and} \bibinfo{person}{Aleš Popovič}.} \bibinfo{year}{2022}\natexlab{}.
\newblock \showarticletitle{Thinking responsibly about responsible AI and ‘the dark side’ of AI}.
\newblock \bibinfo{journal}{\emph{European Journal of Information Systems}} \bibinfo{volume}{31}, \bibinfo{number}{3} (\bibinfo{year}{2022}), \bibinfo{pages}{257--268}.
\newblock
\urldef\tempurl%
\url{https://doi.org/10.1080/0960085X.2022.2026621}
\showDOI{\tempurl}
\showeprint{https://doi.org/10.1080/0960085X.2022.2026621}


\bibitem[Services(2023)]%
        {aws_rekognition_IAM}
\bibfield{author}{\bibinfo{person}{Amazon~Web Services}.} \bibinfo{year}{2023}\natexlab{}.
\newblock \bibinfo{title}{Security and IAM for Amazon Rekognition}.
\newblock
\newblock
\urldef\tempurl%
\url{https://docs.aws.amazon.com/rekognition/latest/dg/security-iam.html}
\showURL{%
\tempurl}


\bibitem[Services(2024a)]%
        {aws_rekognition_regions}
\bibfield{author}{\bibinfo{person}{Amazon~Web Services}.} \bibinfo{year}{2024}\natexlab{a}.
\newblock \bibinfo{title}{Amazon Rekognition Documentation}.
\newblock
\newblock
\urldef\tempurl%
\url{https://docs.aws.amazon.com/general/latest/gr/rekognition.html}
\showURL{%
\tempurl}


\bibitem[Services(2024b)]%
        {aws_compliance_center}
\bibfield{author}{\bibinfo{person}{Amazon~Web Services}.} \bibinfo{year}{2024}\natexlab{b}.
\newblock \bibinfo{booktitle}{\emph{AWS Compliance Center}}.
\newblock
\urldef\tempurl%
\url{https://aws.amazon.com/financial-services/security-compliance/compliance-center/}
\showURL{%
\tempurl}


\bibitem[Services(2024c)]%
        {aws_data_privacy_faq}
\bibfield{author}{\bibinfo{person}{Amazon~Web Services}.} \bibinfo{year}{2024}\natexlab{c}.
\newblock \bibinfo{title}{Data Privacy FAQ}.
\newblock
\newblock
\urldef\tempurl%
\url{https://aws.amazon.com/compliance/data-privacy-faq/}
\showURL{%
\tempurl}


\bibitem[Services(2024d)]%
        {AWS2024DataProtection}
\bibfield{author}{\bibinfo{person}{Amazon~Web Services}.} \bibinfo{year}{2024}\natexlab{d}.
\newblock \bibinfo{booktitle}{\emph{Data Protection}}.
\newblock
\urldef\tempurl%
\url{https://docs.aws.amazon.com/rekognition/latest/dg/data-protection.html}
\showURL{%
\tempurl}


\bibitem[Services(2024e)]%
        {AWS_Face_types}
\bibfield{author}{\bibinfo{person}{Amazon~Web Services}.} \bibinfo{year}{2024}\natexlab{e}.
\newblock \bibinfo{booktitle}{\emph{Face Feature Differences}}.
\newblock
\urldef\tempurl%
\url{https://docs.aws.amazon.com/rekognition/latest/dg/face-feature-differences.html}
\showURL{%
\tempurl}


\bibitem[Services(2024f)]%
        {aws_face_attributes}
\bibfield{author}{\bibinfo{person}{Amazon~Web Services}.} \bibinfo{year}{2024}\natexlab{f}.
\newblock \bibinfo{title}{Guidance on Face Attributes}.
\newblock
\newblock
\urldef\tempurl%
\url{https://docs.aws.amazon.com/rekognition/latest/dg/guidance-face-attributes.html}
\showURL{%
\tempurl}


\bibitem[Services(2024g)]%
        {aws_collections}
\bibfield{author}{\bibinfo{person}{Amazon~Web Services}.} \bibinfo{year}{2024}\natexlab{g}.
\newblock \bibinfo{title}{Managing Collections}.
\newblock
\newblock
\urldef\tempurl%
\url{https://docs.aws.amazon.com/rekognition/latest/dg/collections.html}
\showURL{%
\tempurl}


\bibitem[Services(2024h)]%
        {aws_celebrities}
\bibfield{author}{\bibinfo{person}{Amazon~Web Services}.} \bibinfo{year}{2024}\natexlab{h}.
\newblock \bibinfo{title}{Recognizing Celebrities}.
\newblock
\newblock
\urldef\tempurl%
\url{https://docs.aws.amazon.com/rekognition/latest/dg/celebrities.html}
\showURL{%
\tempurl}


\bibitem[Services(2024i)]%
        {aws_responsible_ai}
\bibfield{author}{\bibinfo{person}{Amazon~Web Services}.} \bibinfo{year}{2024}\natexlab{i}.
\newblock \bibinfo{booktitle}{\emph{Responsible AI}}.
\newblock
\urldef\tempurl%
\url{https://aws.amazon.com/ai/responsible-ai/}
\showURL{%
\tempurl}


\bibitem[Services(2024j)]%
        {aws_face_matching}
\bibfield{author}{\bibinfo{person}{Amazon~Web Services}.} \bibinfo{year}{2024}\natexlab{j}.
\newblock \bibinfo{title}{Responsible AI Practices for Rekognition Face Matching}.
\newblock
\newblock
\urldef\tempurl%
\url{https://docs.aws.amazon.com/ai/responsible-ai/rekognition-face-matching/overview.html}
\showURL{%
\tempurl}


\bibitem[Services(2024k)]%
        {aws_rekognition_vpc}
\bibfield{author}{\bibinfo{person}{Amazon~Web Services}.} \bibinfo{year}{2024}\natexlab{k}.
\newblock \bibinfo{title}{Using Amazon Rekognition with Amazon VPC endpoints}.
\newblock
\newblock
\urldef\tempurl%
\url{https://docs.aws.amazon.com/rekognition/latest/dg/vpc.html}
\showURL{%
\tempurl}


\bibitem[Sivan and Zukarnain(2021)]%
        {Sivan2021}
\bibfield{author}{\bibinfo{person}{Remya Sivan} {and} \bibinfo{person}{Zuriati~A. Zukarnain}.} \bibinfo{year}{2021}\natexlab{}.
\newblock \showarticletitle{Security and Privacy in Cloud-Based E-Health System}.
\newblock \bibinfo{journal}{\emph{Symmetry (Basel)}} \bibinfo{volume}{13}, \bibinfo{number}{5} (\bibinfo{year}{2021}), \bibinfo{pages}{742}.
\newblock
\showISBNx{2073-8994}


\bibitem[Statista(2025)]%
        {statista2025}
\bibfield{author}{\bibinfo{person}{Statista}.} \bibinfo{year}{2025}\natexlab{}.
\newblock \bibinfo{title}{Worldwide Market Share of Leading Cloud Infrastructure Service Providers}.
\newblock
\newblock
\urldef\tempurl%
\url{https://www.statista.com/chart/18819/worldwide-market-share-of-leading-cloud-infrastructure-service-providers/}
\showURL{%
\tempurl}


\bibitem[Union(2020)]%
        {aclu2020}
\bibfield{author}{\bibinfo{person}{American Civil~Liberties Union}.} \bibinfo{year}{2020}\natexlab{}.
\newblock \bibinfo{title}{ACLU Statement on Amazon Face Recognition Moratorium}.
\newblock
\newblock
\urldef\tempurl%
\url{https://www.aclu.org/press-releases/aclu-statement-amazon-face-recognition-moratorium}
\showURL{%
\tempurl}


\bibitem[Union(2021)]%
        {aclu_extended2021}
\bibfield{author}{\bibinfo{person}{American Civil~Liberties Union}.} \bibinfo{year}{2021}\natexlab{}.
\newblock \bibinfo{title}{ACLU Statement on Extended Amazon Face Recognition Moratorium}.
\newblock
\newblock
\urldef\tempurl%
\url{https://www.aclu.org/press-releases/aclu-statement-extended-amazon-face-recognition-moratorium}
\showURL{%
\tempurl}


\bibitem[Wen and Holweg(2024)]%
        {Wen&Holweg2024}
\bibfield{author}{\bibinfo{person}{Yuni Wen} {and} \bibinfo{person}{Matthias Holweg}.} \bibinfo{year}{2024}\natexlab{}.
\newblock \showarticletitle{A phenomenological perspective on AI ethical failures: The case of facial recognition technology}.
\newblock \bibinfo{journal}{\emph{AI \& society}} \bibinfo{volume}{39}, \bibinfo{number}{4} (\bibinfo{year}{2024}), \bibinfo{pages}{1929--1946}.
\newblock
\showISBNx{0951-5666}


\bibitem[Woolf(2018)]%
        {aws_gdpr}
\bibfield{author}{\bibinfo{person}{Chad Woolf}.} \bibinfo{year}{2018}\natexlab{}.
\newblock \bibinfo{booktitle}{\emph{The AWS Shared Responsibility Model and GDPR}}.
\newblock
\urldef\tempurl%
\url{https://aws.amazon.com/blogs/security/the-aws-shared-responsibility-model-and-gdpr/}
\showURL{%
\tempurl}


\bibitem[Xu et~al\mbox{.}(2021)]%
        {Xu2021}
\bibfield{author}{\bibinfo{person}{Feng~Z. Xu}, \bibinfo{person}{Yun Zhang}, \bibinfo{person}{Tingting Zhang}, {and} \bibinfo{person}{Jing Wang}.} \bibinfo{year}{2021}\natexlab{}.
\newblock \showarticletitle{Facial recognition check-in services at hotels}.
\newblock \bibinfo{journal}{\emph{Journal of hospitality marketing \& management}} \bibinfo{volume}{30}, \bibinfo{number}{3} (\bibinfo{year}{2021}), \bibinfo{pages}{373--393}.
\newblock
\showISBNx{1936-8623}


\bibitem[Yang et~al\mbox{.}(2021)]%
        {YangCloudFRT}
\bibfield{author}{\bibinfo{person}{Chia-En Yang}, \bibinfo{person}{Yang-Ting Shen}, {and} \bibinfo{person}{Shih-Hao Liao}.} \bibinfo{year}{2021}\natexlab{}.
\newblock \showarticletitle{SyncBIM: The Decision-Making BIM-Based Cloud Platform with Real-time Facial Recognition and Data Visualization}.
\newblock \bibinfo{journal}{\emph{Advances in Science, Technology and Engineering Systems Journal (ASTESJ)}} \bibinfo{volume}{6}, \bibinfo{number}{5} (\bibinfo{year}{2021}), \bibinfo{pages}{16--22}.
\newblock
\showISBNx{2415-6698}


\bibitem[Yang(2021)]%
        {Yang2021}
\bibfield{author}{\bibinfo{person}{Qiang Yang}.} \bibinfo{year}{2021}\natexlab{}.
\newblock \showarticletitle{Toward Responsible AI: An Overview of Federated Learning for User-centered Privacy-preserving Computing}.
\newblock \bibinfo{journal}{\emph{ACM transactions on interactive intelligent systems}} \bibinfo{volume}{11}, \bibinfo{number}{3-4} (\bibinfo{year}{2021}), \bibinfo{pages}{1--22}.
\newblock
\showISBNx{2160-6455}


\bibitem[Yang et~al\mbox{.}(2022)]%
        {YangCloud2021}
\bibfield{author}{\bibinfo{person}{Tao Yang}, \bibinfo{person}{Yuhang Zhang}, \bibinfo{person}{Jie Sun}, {and} \bibinfo{person}{Xun Wang}.} \bibinfo{year}{2022}\natexlab{}.
\newblock \showarticletitle{Privacy Enhanced Cloud-Based Facial Recognition}.
\newblock \bibinfo{journal}{\emph{Neural processing letters}} \bibinfo{volume}{54}, \bibinfo{number}{4} (\bibinfo{year}{2022}), \bibinfo{pages}{2717--2725}.
\newblock
\showISBNx{1370-4621}


\end{thebibliography}

\appendix

\end{document}